\documentclass[utf8]{frontiersFPHY} 

\usepackage{microtype,subcaption}
\usepackage[onehalfspacing]{setspace}
\usepackage{placeins}

\def\keyFont{\fontsize{8}{11}\helveticabold }
\def\firstAuthorLast{Ottolina {et~al.}} 
\def\Authors{Ottolina R.\,$^{1,*}$, Falomo R.\,$^{2}$, Treves A.\,$^{1}$, Uslenghi M.\,$^{3}$, Kotilainen J.K.\,$^{4}$, Scarpa R.\,$^{5}$ and Farina E.P.\,$^{6}$ }
\def\Address{$^{1}$ DISAT-Universit\`{a} degli Studi dell'Insubria, Como, Italy \\
$^{2}$ INAF-Osservatorio astronomico di Padova, Italy \\
$^{3}$ INAF/IASF -Milano, Italy \\
$^{4}$ Finnish Centre for Astronomy with ESO (FINCA), Finland \\
$^{5}$ Istituto de astrofisica de Canarias, Spain \\
$^{6}$ Max-Planck-Institut f\"{u}r Astronomie, Heidelberg, Germany}
\def\corrAuthor{Via Valleggio 11, COMO 22100 ITALY }
\def\corrEmail{r.ottolina@uninsubria.it}

\begin{document}
\onecolumn
\firstpage{1}

\title[Circum-galactic medium in quasar haloes]{Circum-galactic medium in the halo of quasars} 

\author[\firstAuthorLast ]{\Authors} 
\address{} 
\correspondance{} 

\extraAuth{}

\maketitle

\begin{abstract}

\section{}
The properties of circum-galactic gas in the halo of quasar host galaxies are investigated analyzing Mg II 2800 and C IV 1540 absorption-line systems along the line of 
sight close to quasars. We used optical spectroscopy of closely aligned pairs of quasars (projected distance $\leq$ 200 kpc, but at very different redshift) obtained at 
the VLT and  Gran Telescopio Canarias to investigate the distribution of the absorbing gas for a sample of  quasars at z$\sim$1. Absorption systems of 
EW $\geq$ 0.3 $\rm{\AA}$ associated with the foreground quasars are revealed up to 200 kpc from the centre of the host galaxy, showing that the structure of the absorbing gas is patchy with a covering fraction 
quickly decreasing beyond 100 kpc. In this contribution we use optical and near-IR images obtained at VLT to investigate the relations between the properties of the circum-galactic medium of the host galaxies 
and of the large scale galaxy environments of the foreground quasars.
\tiny
 \keyFont{ \section{Keywords:} quasar, quasar environment, quasar pair, quasar spectra, galaxy around quasar} 
\end{abstract}

\section{Introduction}
The standard model for the origin of the extreme luminosity of quasars considers that a supermassive black hole shines as a quasar when intense mass inflow 
takes place, possibly as a consequence of tidal forces in dissipative events (e.g., \cite{art:Dimatteo2005}). In this scenario, the circum-galactic medium
 of quasar host galaxies is expected to be populated by streams, cool gas clouds and tidal debris, as commonly observed in interacting galaxies 
(e.g. \cite{art:Sulentic2001},\cite{art:Cortese2006}). Moreover the gas of the circum-galactic medium could be metal enriched by supernova-driven winds 
triggered by starbursts 
events associated to the mergers or by quasar-driven outflows of gas (e.g., \cite{art:Steidel2010}, \cite{art:Shen2012}). 

\begin{figure}[htpb]
\centering\includegraphics[width=\linewidth]{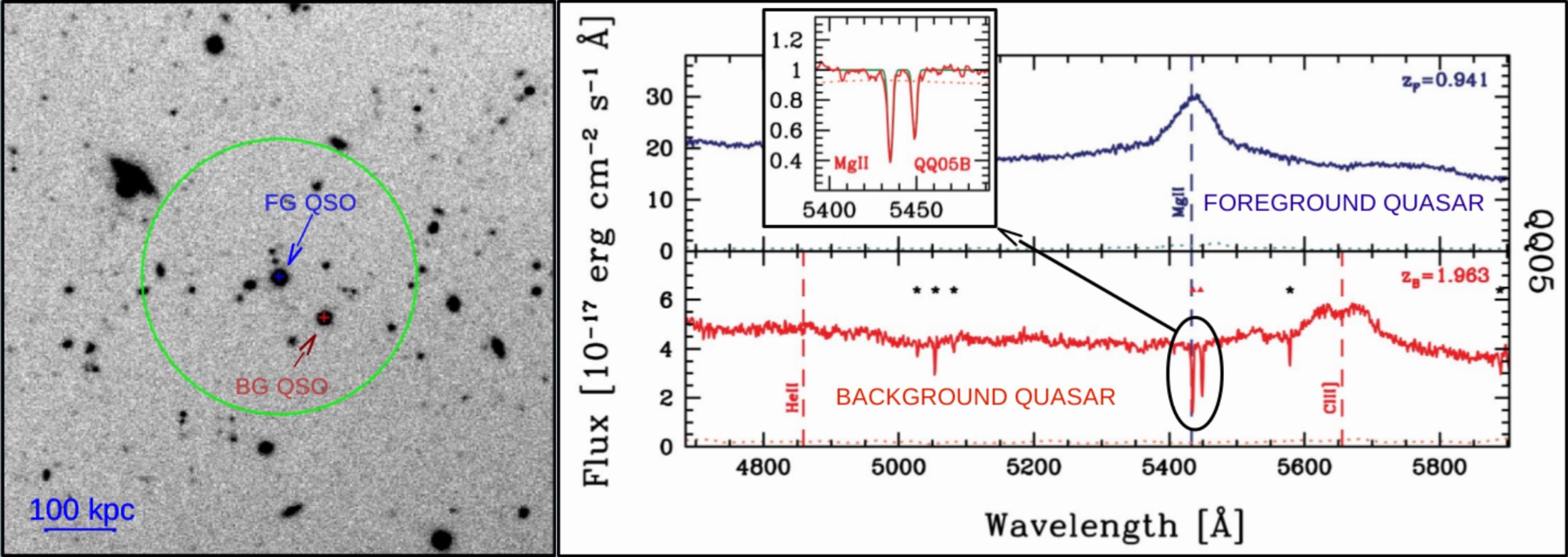}
\caption{Image of projected quasar pair Q0059-2702. Blue and red arrows indicate the foreground quasar and the background one. Green circle shows the projected distance of 200 kpc from the foreground quasar.
\emph{Right}: VLT spectra of quasar pair Q0059-2702. The blue and red solid lines refer to foreground quasar and background quasar, respectively. The insert shows the zoom of Mg II absorptions 
at the same redshift of foreground quasar.}\label{fig:1}
\end{figure}

One of the effective ways to study the circum-galactic medium of galaxies at high redshift is to investigate the absorption features that they imprint in the spectra of quasars. 
In particular, projected quasar pairs (figure \ref{fig:1} left) are ideal observational tools for this purpose, since the light of the very bright source in the background (z$\equiv$ $\rm{z}_{\rm{B}}$) goes through the extended halo 
of the foreground (z$\equiv$ $\rm{z}_{\rm{F}}$ $<$ $\rm{z}_{\rm{B}}$) object (e.g. \cite{art:Hennawi2006}, \cite{art:Farina2013}). This can be evidenced by absorption lines at the foreground redshift: an example is reported in figure \ref{fig:1} right.

\begin{figure}[htpb]
 \centering\includegraphics[width=\linewidth]{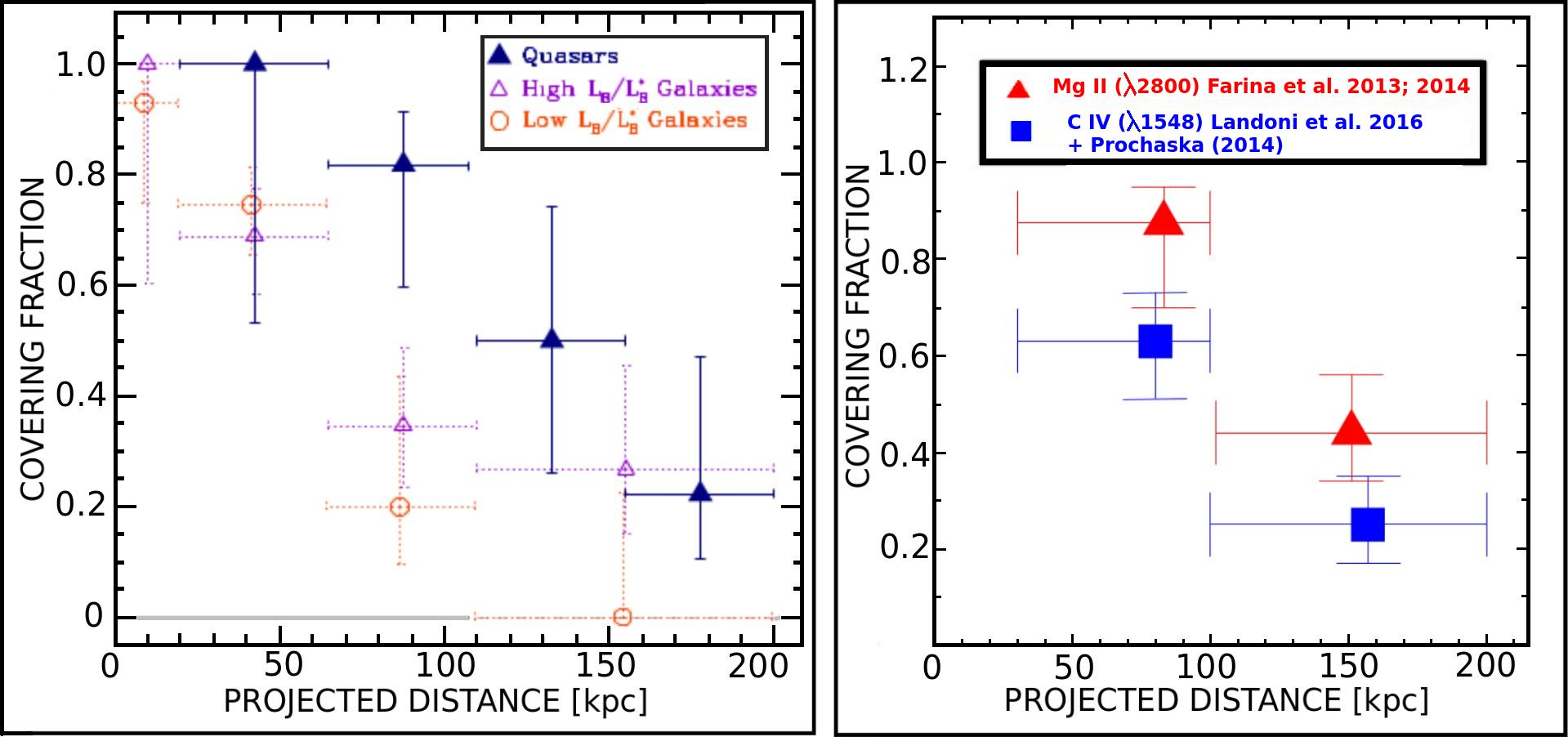}
\caption{Comparison between quasars \cite{art:Farina2014} and galaxies \cite{art:Nielsen2013} of covering fraction of transversal absorption system of Mg II as a function of the projected 
distance. \emph{Right}: Comparison of covering fraction of quasar halo for Mg II (red triangles) and C IV (blue squares) absorption lines. 
The covering fraction  of Mg II is systematically higher than that of C IV.}\label{fig:2}
\end{figure}

In our previous works we studied 49 quasar pairs (\cite{art:Farina2013},\cite{art:Farina2014},\cite{art:Landoni2016}).
We used the optical spectroscopy of close pairs (projected distance $\leq$ 200 kpc) obtained at the ESO-VLT and Gran Telescopio Canarias (GTC) to investigate the distribution of the absorbing gas at 100-200 kpc projected distance 
from the quasar studying the presence of Mg II or C IV absorption lines at the redshift of foreground quasar. 
In order to characterize the structure of circum-galactic medium of the foreground quasar host galaxy we estimated the covering fraction of Mg II or C IV as a function of the
projected distance. We assumed a threshold in equivalent width of 0.3 $\rm{\AA}$, then we subdivided the projected distance in bins. For each bin we computed the covering 
fraction as the ratio between the number of systems with Mg II or C IV absorption lines greater than the threshold and the total number of observed systems. Our previous results (\cite{art:Farina2013},\cite{art:Farina2014},\cite{art:Landoni2016}) indicate that 22 spectra exhibit absorption lines of foreground 
quasar in the background quasar: 15 Mg II and 7 C IV. We found that the detected Mg II absorption systems of EW $>$ 0.3 $\rm{\AA}$ associated with the foreground quasars are revealed up to $\sim$200 kpc from the centre of the host galaxy. 
The structure of absorbing gas is patchy with covering fraction of the gas that quickly decreases beyond 100 kpc. This is illustrated in figure \ref{fig:2} left. A similar behavior is present in absorption systems with C IV doublet (figure \ref{fig:2} right).

In this paper we analyze optical and NIR images of foreground quasars in order to investigated their closed environments and their host galaxies.  

\section{Sample and observations}
The selection procedure of our quasar pair projected sample is reported in detail in \cite{art:Farina2013} and  \cite{art:Farina2014}. 
Here we summarize briefly the main criteria of our choice:
i) visibility from telescope location;
ii) foreground quasar redshift must be selected in order that Mg II doublet falls in GRISM wavelength range;
iii) projected distance at foreground redshift $\leq$ 200 kpc;
iv) line-of-sight velocity difference $\geq$ 5000 km s$^{-1}$ to avoid physical pairs;
v) background quasar must be brighter than $\rm{m}_{V}\sim21$ to ensure good spectra signal-to-noise ratio.

We acquired optical images at FORS2@VLT with the I BESS filter for the object with z$\leq$1 and the Z GUNN filter otherwise, and near-IR images at HAWK-I@VLT with 
H filter for a total of 24 quasar projected pairs (4 targets have only optical images, 8 ones have only NIR images and 12 objects have images in both bands.).
With this configuration we explore galaxies at the redshift of the target in the B band in the rest frame. 
For a subsample of object reported in table \ref{tab:1} we present here 8 deep high quality I-band images (seeing $\sim$0.5 arcsec).  
\begin{table}[h] \small
 \begin{center}
 \begin{tabular}{lclcccc}
QSO$_{\rm{B}}$             &  z$_{\rm{B}}$ &        QSO$_{\rm{F}}$    & z$_{\rm{F}}$ & pd [kpc]  &  Mg II & ovdens\\
(1) & (2)&(3)&(4)&(5)&(6)&(7)\\
\hline        
\ \\
 J003823.6-291259    & 2.699 & J003823.74-291311.8 & 0.793 &  91 & no  & no \\     
 LQAC 015-026 011    & 1.963 & J010204.12-264600.0 & 0.941 &  84 & yes & no \\     
 J013500.09-004113.4 & 1.259 & J013458.77-004129.0 & 1.003 & 176 & yes & yes\\    
 J014630.95+001531.6 & 1.019 & J014630.14+001521.3 & 0.923 & 125 & yes & yes\\    
 J021553.71+010953.9 & 2.215 & J021552.53+011000.1 & 0.875 & 145 & no  & yes\\     
 J022158.83-001052.5 & 3.213 & J022158.77-001044.3 & 1.036 &  66 & yes & yes\\ 	 
 $[$HB89$]$ 2225-403 & 2.398 & J222850.49-400825.7 & 0.931 &  78 & yes & no \\      
 J225902.37+003221.7 & 1.456 & J225902.87+003243.7 & 0.868 & 183 & no  & no \\   
\hline
\end{tabular}
\caption{List of targets observed in I band. Column (1) and (3) give the name of background and foreground quasars respectively while (2) and (4) give their redshifts.
Column (5) reports the projected distances at the redshift of the foreground quasar. Column (6) and (7) are labels for the presence of Mg II absorption lines and of an overdensity of 
galaxies around the foreground quasar.}\label{tab:1}
\end{center}
\end{table}
\section{Analysis}
We performed the analysis of optical images in order to characterize the close environment of foreground quasar. We used the software SEXTRACTOR \cite{art:Bertin1996}
to identify all objects in the frame over a given magnitude limit and to distinguish galaxy-like objects from star-like ones (galaxies have CLASS$\_$STAR $<$ 0.5 and stars have CLASS$\_$STAR $>$ 0.5).
Then we evaluated the overdensity of galaxies around the foreground quasar calculating the ratio of number of galaxies per arcminute square to background estimated at distances greater than 500 kpc.

The near-IR images have been analyzed using the software package AIDA (Astronomical Image Decomposition Analysis, \cite{art:Uslenghi2008}).
From this analysis of the near-IR images we are able to characterize the properties of the foreground quasars host galaxy via 2-d model fitting, 
assuming that they are the result of the superposition of two components: the nucleus, described by the local PSF, and the host galaxy, modeled by a Sersic 
function convolved with the proper PSF. 

\section{Preliminary results}
\begin{figure}[htpb]
 \centering\includegraphics[width=\linewidth]{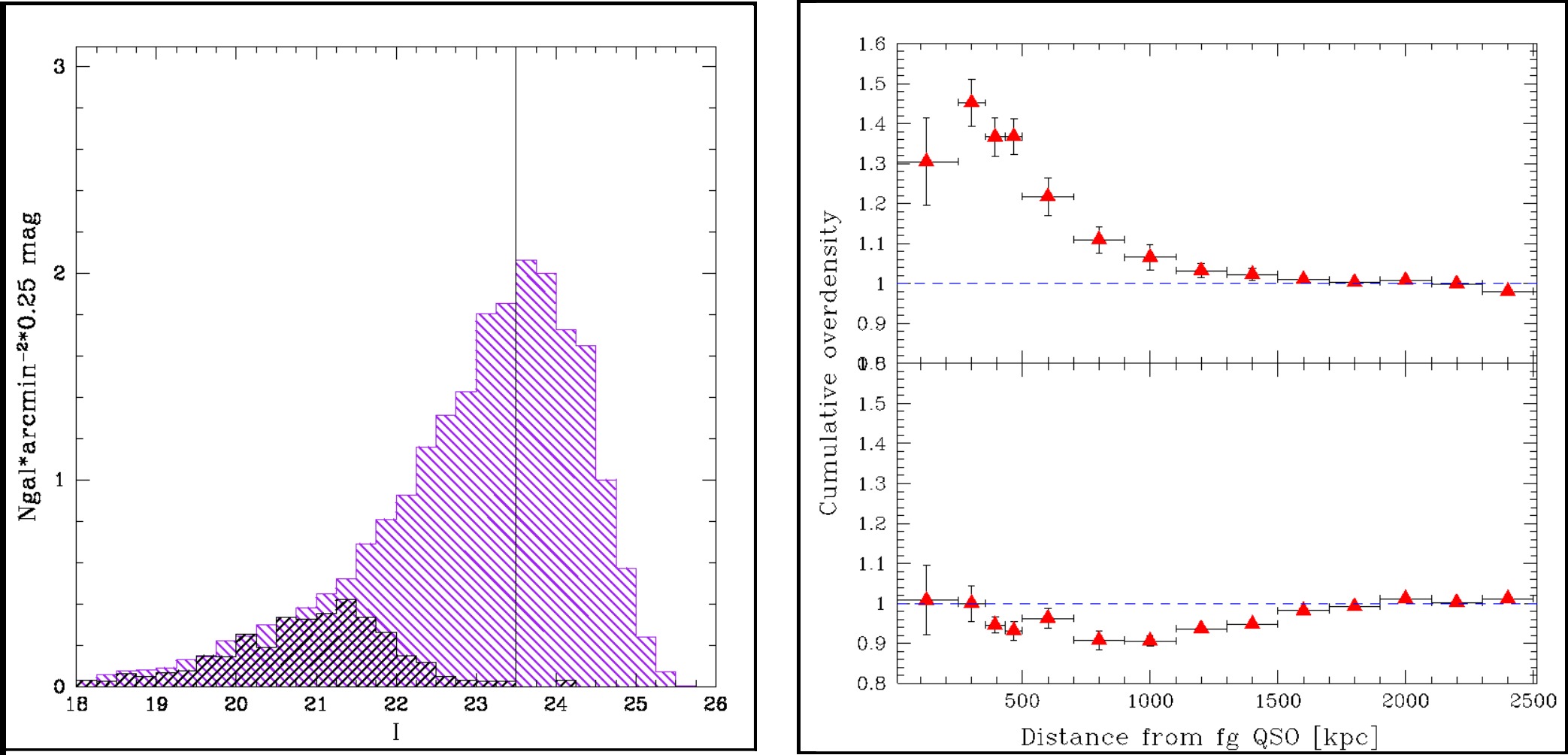}
\caption{\emph{Left}: Average surface number density of galaxies vs I magnitude of 8 quasar pair fields obtained at VLT FORS2.
The vertical line marks the adopted threshold magnitude for the environment study. For comparison the black histogram shows the similar distribution based on the SDSS analysis of 5 fields.
\emph{Right}: Average cumulative overdensity of galaxies around  quasars (see table \ref{tab:1}). Only half of quasars exhibits a clear galaxy 
overdensity (upper panel) while the other half does not show any overdensity (lower panel).}\label{fig:3}
\end{figure}

Based on our deep optical images of quasars we are able to characterize the galaxy environment up to I$\sim$ 23.5  which is more than 2 magnitudes deeper than SDSS images (see figure \ref{fig:3} left).
This allows to investigate the galaxy environment down to about 3 magnitude fainter than M$^{*}$. We find that for 4 cases there is a clear galaxy overdensity around the foreground quasar while in the another 4 cases there is no evidence that quasars live in a group of galaxies
(see figure \ref{fig:3} and table \ref{tab:1}). In 3 quasars that exhibit overdensity there is also a detection of Mg II absorption systems at the same redshift of foreground quasar in the circum-galactic medium.
In the cases of no galaxy overdensity Mg II absorption lines are detected in 2 objects. The small sample, investigated till now, does not permit us to draw firm conclusions on the relationship between galaxy environments and presence 
of cold gas in the intergalactic medium. We are completing the 
analysis of the full sample and extending it with other targets from ongoing observations at GTC.

Till now we analyzed 11 out of the 20 foreground quasars with NIR images. For 9 ($\sim$ 80\%) the host galaxy is well resolved.
In one case the detection of the host galaxy is marginal and only for one image no evidence of the host galaxy is found. 
The rest frame absolute magnitude in I band of the resolved host galaxies ranges from -22.5 to -24.9, with a median value of -23.7. 
These results are comparable to those reference to other quasars at similar redshift \cite{art:Sanghvi2014}. 

\section{Acknowledgment}
EPF acknowledge funding through the ERC grant `Cosmic Dawn'.

\FloatBarrier
\bibliographystyle{frontiersinHLTH&FPHY} 
\bibliography{biblio_Ottolina_PADOVA}

\end{document}